\documentstyle[12pt]{article}
\def\rZ{{\rm I\!R}}

\def\gZ{{\rm Z\!\!Z}}
\begin{document}

\centerline{\bf Remarks on `` COLORING RANDOM TRIANGULATION''}

\baselineskip=16pt

\vspace*{0.6cm}
\centerline{\footnotesize S.BALASKA\footnote{On leave of absence from Laboratoire de Physique Th\'eorique, Universit\'e d'Oran, 31100 Es-S\'enia, Algeria, E-mail: balaska@physik.uni-kl.de}  and  
W. R\"UHL\footnote{E-mail:ruehl@physik.uni-kl.de}}
\baselineskip=13pt
\centerline{\footnotesize\it Department of Physics, University of Kaiserslautern, P.O.Box 3049}
\centerline{\footnotesize\it 67653 Kaiserslautern, Germany}
\vspace*{0.8cm}
\vspace*{3cm}

\begin{abstract}
We transform the two-matrix model, studied by P.Di Francesco and al. in \cite{1}, into a normal one-matrix model by identifying a ``formal'' integral used by these authors as a proper integral. We show also, using their method, that the results obtained f
or the resolvent and the density are not reliable.
\end{abstract}
\vspace*{3cm}
\section{THE MODEL}
In a recent paper P. Di Francesco, B. Eynard and E. Guitter \cite{1} discuss
a model of two $n \times n$ hermitean matrices $M$ and $R$ with a partition 
function
\begin{equation}
Z(p,q,g;N)=\int dM dR  \exp\{-N Tr[p \log(1-M) + q \log(1-R) + gMR ]\}
\label{(1)}
\end{equation}
where $g$ is later replaced by
\begin{equation}
g = \frac{1}{t}
\label{(2)}
\end{equation} 
and the strong coupling limit is of primary interest: $t \rightarrow 0$.
Applying the Itzykson-Zuber integral identity, the partition function is 
transformed into an integral over the eigenvalues $\{ m_i\}_{i=1,..n}$ and 
$\{r_i\}_{i=1,..n}$ of $M$ and $R$ respectively.
\begin{eqnarray}\label{(3)}
Z(p,q,g;N) \sim \int \Delta(m) \Delta(r) \prod_{i=1}^{n} [ \exp \{-N [p \log(1-m_i) \\ \nonumber
+ q \log(1-r_i) + g m_i r_i]\} dr_i dm_i]
\end{eqnarray}
where $\Delta$ is the Vandermonde determinant.

However, an integral such as
\begin{eqnarray}
\int_{\rZ^2} dx dy  \exp \{-N[p\log(1-x) + q\log(1-y) + gxy] \} p_n(x)\tilde{p}_m(y)
\label{(4)}
\end{eqnarray}
where $p_n (\tilde{p}_m)$ are polynomials of degree $n(m)$ does not exist.

In later parts of the paper the authors want to ascribe a meaning to such integrals by the ``formal integral'' (see eq. (4.2))
\begin{eqnarray}
\int_{\rZ^2} dxdy \,\,x^{\alpha} y^{\beta} \exp[-\frac{N}{t} xy] &=& \alpha! \ \delta_{\alpha \beta}\,\,\, (\frac{t}{N})^{\alpha +1} \\ \nonumber
(\alpha,\beta \in \gZ_+)&& 
\label{(5)}
\end{eqnarray} 
Special cases of such ``formal integral'' are already used earlier in the text.
It is, however, easy to see that the ``formal integral'' is a proper integral
\begin{eqnarray}
2\int_{\rZ^2} dxdy \,\,w^{\alpha} \bar{w}^{\beta} \exp[-\frac{N}{t} w\bar{w}] &=& \alpha! \ \delta_{\alpha \beta}\,\,\, (\frac{t}{N})^{\alpha +1} \\ \nonumber
(w=x+iy &,& \bar{w}=x-iy)
\label{(6)}
\end{eqnarray} 
It is also easy to go back and replace all ``formal integrals'' by proper integrals, then for (\ref{(3)}) we obtain
\begin{eqnarray}
Z(p,q,g;N) \sim \int \mid\Delta(w)\mid^2\prod_{i-1}^{n} [ \exp \{-N [p \log(1-w_i) \\ \nonumber
+ q \log(1-\bar{w}_i) + g w_i \bar{w}_i]\} dx_i dy_i]
\label{(7)}
\end{eqnarray}
where we have to restrict the $w_i$ to, say 
\begin{eqnarray} 
0 \leq \mid w_i \mid \leq 1 
\label{8}
\end{eqnarray}
and
\begin{eqnarray}
Re(Np)<1 , Re(Nq)<1
\label{(9)}
\end{eqnarray}
if we want to give an analytic meaning to Z in the variable $t=g^{-1}$, but we 
can integrate over the whole complex plane if we are interested only in the formal power expansion in $t$ for $t \rightarrow 0$. Finally, without referring to the Itzykson-Zuber formula we can integrate over $U(n)$ to get
\begin{equation}
Z(p,q,g;N)= 2^n \int dM dM^{\dagger} \exp\{-N Tr[p \log(1-M) + q \log(1-M^{\dagger}) + gMM^{\dagger} ]\}
\label{(10)}
\end{equation}
where $M$ is a normal matrix
\begin{eqnarray}
[M,M^{\dagger}]=0
\label{(11)}
\end{eqnarray}
Thus the model actually evaluated is not a hermitean two-matrix model but a normal one-matrix model.

\section{THE SADDLE POINT EQUATION}
The partition function is calculated in the limit $N\rightarrow \infty$ so that $n=zN$ and $z$ is kept fixed. In this limit the partition function is approximated by (see eq.(4.20) of \cite{1})
\begin{eqnarray}
Z = N^n \int_0^{\infty} d\alpha_i...d\alpha_n \exp\{N^2 S(\alpha_1..\alpha_n,p,q,z)\}
\label{(12)}
\end{eqnarray}
with
\begin{eqnarray} 
S=\frac{1}{N}\sum_{i=1}^{n}&\Bigl[&(\alpha_i+p-z)(\log(\alpha_i+p-z)-1) 
\nonumber \\
&+& (\alpha_i+q-z)(\log(\alpha_i+q-z)-1)-2\alpha_i(\log(\alpha_i)-1)\nonumber \\
&+&\frac{1}{N} \log (\gamma(\alpha_i N +1 ,\frac{N}{t})) + \alpha_i\log(\frac{t}{N})\Bigr]  \nonumber \\
&+&\frac{1}{N^2} \sum_{i \neq j} \log(\mid \alpha_i-\alpha_j\mid)  \nonumber \\
&-& \int_0^z ds [ (p-s)(\log(p-s)-1) + (q-s)(\log(q-s)-1)   \nonumber \\ 
&&+ s(\log(s)-1) + s\log(t)] 
\label{13}
\end{eqnarray}
where $\gamma(\alpha,\xi)$ is the incomplete $\gamma$-function \cite{2}. In the limit $N \rightarrow \infty$ the term containing the incomplete $\gamma$-function is calculated as
\begin{equation}
\lim_{N \to \infty} \Bigl\{ -\frac{t}{N} \log(\gamma(\xi \frac{N}{t}+1,\frac{N}{t}))-\xi \log(\frac{t}{N}) \Bigr\} = \Theta(\xi-1) + \xi(1-\log(\xi))\Theta(1-\xi)
\label{(14)}
\end{equation}
where $\xi=\alpha t$ and so that the derivative of the r.h.s is continuous at $\xi=1$.
The saddle point equation is then
\begin{eqnarray}\label{(15)}
2P\int d\beta \frac{\rho(\beta)}{\beta-\alpha} &=& \Theta(1-\alpha t) \log \Bigl( t \frac{(\alpha+p-z)(\alpha+q-z)}{\alpha} \Bigr) \nonumber \\
&+& \Theta(\alpha t -1) \log \Bigl( \frac{(\alpha+p-z)(\alpha+q-z)}{\alpha^2}\Bigr)
\end{eqnarray}
valid for $\alpha$ in the support of the density $\rho$ and with the 
normalisation
\begin{equation}
\int d\beta \rho(\beta) = z
\label{(16)}
\end{equation}
We mention that the free energy also has an additional term compared with the 
expression obtained in the equation (4.29) of \cite{1}
\begin{equation}
t\partial_t f = \int_0^{\infty} d\beta \beta \rho(\beta) -\frac{1}{2} z^2 + \frac{1}{t} \int_{t^{-1}}^{\infty} d\beta \rho(\beta)
\label{(17)}
\end{equation}
We solve the saddle point equation by making the ansatz
\begin{eqnarray}
\rho(\alpha) &=& \rho_1(\alpha) + \rho_2(\alpha) \label{(18)} \\
\rho_i(\alpha) \geq 0 \,\,\,\,\, &,& \,\,\,\, \int d\beta \rho_i(\beta)=z_i \label{(19)}\\
z_1 + z_2 &=& z  \label{(20)}
\end{eqnarray}
and 
\begin{eqnarray}
supp (\rho_1) &=& < \gamma_1,\gamma_2> \label{(21)}\\
supp (\rho_2) &=& < \gamma_3,\gamma_4> \label{(22)}
\end{eqnarray}
In order to solve (\ref{(15)}) we have to make sure that
\begin{equation}
\gamma_2 \leq \frac{1}{t} \leq \gamma_3
\label{(23)}
\end{equation}
Next we introduce the resolvent
\begin{equation}
\omega_i(\alpha) = \int d\beta \frac{\rho_i(\beta)}{\alpha-\beta}
\label{(24)}
\end{equation}
so that (\ref{(15)}) induces
\begin{eqnarray}
\omega_1(\alpha+i0) + \omega_1(\alpha-i0) = - \log \Bigl( t \frac{(\alpha+p-z)(\alpha+q-z)}{\alpha} \Bigr) \label{(25)} \\
\omega_2(\alpha+i0) + \omega_2(\alpha-i0) = -\log \Bigl( \frac{(\alpha+p-z)(\alpha+q-z)}{\alpha^2}\Bigr) \label{(26)}
\end{eqnarray}
and
\begin{equation}
\lim_{\alpha \to \infty} \alpha \omega_i(\alpha) = \int d\beta \rho_i(\beta) = z_i
\label{(27)}
\end{equation}
We shall use the method of \cite{1}, and show that their result for $\omega_1, \rho_1$ is not reliable, in spite of the fact that the saddle point equation (\ref{(25)}) is the same as the one obtained in \cite{1}. For simplicity we set 
$z_1=z,z_2=0$ from now on.

We introduce two parameters $r, \delta$ and three hyperbolic angles (assumed to be all $\geq 0$)$\Phi_1,\Phi_2,\Phi_3$ by
\begin{eqnarray}
\alpha &=& z-r-2\delta \cosh(\Phi) \label{(28)}\\
p&=&r+2\delta \cosh(\Phi_1)\label{(29)}\\
q&=&r+2\delta \cosh(\Phi_2)\label{(30)}\\ 
z&=&r+2\delta \cosh(\Phi_3)\label{(31)}
\end{eqnarray}
Then
\begin{equation}
\frac{t}{\alpha}(\alpha+p-z)(\alpha+q-z)=\delta t \frac{4(T_1^2-T^2)(T_2^2-T^2)
(1-T_3^2)}{(1-T_1^2)(1-T_2^2)(1-T^2)(T_3^2-T^2)}
\label{32}
\end{equation}  
where
\begin{eqnarray}
T&=&\tanh(\frac{\Phi}{2}) \label{33} \\
T_i&=&\tanh(\frac{\Phi_i}{2}) \geq 0 \label{34}
\end{eqnarray}
with
\begin{eqnarray*}
\gamma_1=z-r-2\delta \,\,\, , \,\,\, \gamma_2=z-r+2\delta
\end{eqnarray*}
Along the cut $<\gamma_1,\gamma_2>$ the parameter $\Phi$ is assumed to vary as 
$+i\varphi$ $(0 \leq \varphi \leq \pi)$ 
for $Im(\alpha) \searrow 0$ and as $-i\varphi$ for $Im(\alpha) \nearrow 0$. 
On the real axis outside the cut, $\Phi$ is necessarily so that
\begin{equation}
T= \tanh (\frac{\Phi}{2}) \rightarrow -1 \,\,\,\,\,\mbox{for} \,\,\,\,\,\alpha \rightarrow \pm\infty
\label{35}
\end{equation}
Taking into account (\ref{34}) and (\ref{35}) only one ansatz for 
$\omega_1(\alpha)$ is possible which has only the cut at $<\gamma_1,\gamma_2>$, 
namely
\begin{equation}
\omega_1(\alpha)= - \log \Bigl(\frac{2(T_1-T)(T_2-T)(1+T_3)}{(1-T)(1+T_1)(1+T_2)
(T_3-T)} \Bigr)
\label{36}
\end{equation}
provided we take 
\begin{eqnarray}
\delta t &=& \frac{(1-T_1)(1-T_2)(1+T_3)}{(1+T_1)(1+T_2)(1-T_3)} \nonumber \\
&=& u_1 u_2 u_3^{-1}
\label{37}
\end{eqnarray}
where
\begin{eqnarray*}
u_i &=& \frac{1-T_i}{1+T_i} = e^{-\Phi_i}\\
u&=&\frac{1-T}{1+T} = e^{-\Phi}
\end{eqnarray*}
Finally we obtain for the density
\begin{equation}
\rho_1(\alpha) = \frac{1}{2\pi}\Bigl(\varphi -2(\Psi_1(\varphi)+\Psi_2(\varphi)-\Psi_3(\varphi)) \Bigr) 
\label{38}
\end{equation}
where
\begin{eqnarray*}
\Psi(\alpha) = \arctan (\frac{\tan(\frac{\varphi}{2})}{T_i})
\end{eqnarray*}

We can show that
\begin{eqnarray}
z&=& \int \rho_1(\beta) d\beta \nonumber \\
&=& \int_0^{\pi} d\varphi 2 \delta \sin(\varphi) \rho_1(\beta(\varphi)) \nonumber \\
&=& \delta [u_3-u_1-u_2] \nonumber \\
&=& \lim_{\alpha \to +\infty} \alpha \omega_1(\alpha)
\label{43}
\end{eqnarray}
as desired.

Positivity of the density function (\ref{38}) is achieved if and 
only if
\begin{equation}
T_1+T_2-T_3-1 \geq 0
\label{44}
\end{equation}
This constraint follows from the observation that the neighborhood of $\pi$ in the variable $\varphi$ is critical, i.e. negative values of $\rho_1$ appear here 
first. Namely, if we write
\begin{equation} \label{45}
\frac{\varphi}{2} = \frac{\pi}{2} - \chi
\end{equation}
we have then
\begin{equation}
\Psi_i(\varphi) = \frac{\pi}{2}-T_i \chi + O(\chi^3)
\label{46}
\end{equation}
and consequently
\begin{equation}
\rho_1(\alpha) = \frac{1}{\pi}(T_1+T_2-T_3-1) \chi + O(\chi^3)
\label{47}
\end{equation}

\end{document}